\documentclass{ws-procs9x6}
\usepackage{graphicx}
\usepackage{epsfig}

\begin{document}
%
\def\slasha#1{#1\hskip-0.65em /}  
\def\slashb#1{#1\hskip-1.3em /}   
\def\slashc#1{#1\hskip-.4em /}
%
\def \pb        {{\rm \, pb}}
\def \fb        {{\rm \, fb}}
\def \ipb       {{\rm \, pb^{-1}}}
\def \eV        {{\rm \,  eV}}
\def \keV       {{\rm \, keV}}
\def \MeV       {{\rm \, MeV}}
\def \GeV       {{\rm \, GeV}}
\def \TeV       {{\rm \, TeV}}
\def \MHz       {{\rm \, MHz}}
\def \mrad      {{\rm \, mrad}}
\def \TeVc      {\TeV/c}
\def \TeVcc     {\TeV/c^2}
\def \GeVc      {\GeV/c}
\def \GeVcc     {\GeV/c^2}
\def \MeVc      {\MeV/c}
\def \MeVcc     {\MeV/c^2}
%
\def\ga{\mathrel{\raise.3ex\hbox{$>$\kern-.75em\lower1ex\hbox{$\sim$}}}}
\def\la{\mathrel{\raise.3ex\hbox{$<$\kern-.75em\lower1ex\hbox{$\sim$}}}}
\newcommand{\ckm}{$\checkmark$}
%
\newcommand {\slashed}[1] { \mbox{\rlap{\hbox{/}} #1 }}
\newcommand {\onehalf}    {\raisebox{0.1ex}{${\frac{1}{2}}$}}
\newcommand {\fivethirds} {\raisebox{0.1ex}{${\frac{5}{3}}$}}
\newcommand {\OR}         {{\tt OR}\,}
\newcommand {\BR}         {{\rm BR}\,}
\newcommand {\rts}        {\sqrt{s}}
\newcommand {\lumi}       {\mathcal{L}}
\newcommand {\Lumi}       {\int\lumi\mathrm{d}t}
\newcommand {\gradi}    {^\circ}
\newcommand {\de}         {\partial}
\newcommand {\um}         {\mu \rm m}
\newcommand {\nm}         {\rm   nm}
\newcommand {\us}         {  \mu \rm s}
\newcommand {\cm}         {\rm   cm}
\newcommand {\mm}         {\rm   mm}
\newcommand {\m}          {\rm   m}
\newcommand {\km}         {\rm   km}
\newcommand {\V}          {\rm   V}
\newcommand {\T}          {\rm   T}
\newcommand {\kV}         {\rm   kV}
\newcommand {\kVm}        {\rm   kV/m} 
\newcommand {\MVm}        {\rm   MV/m} 
\newcommand {\ns}         {\rm   ns} 
%
\newcommand {\gws}        {\mathrm{SU(2)_L \otimes U(1)_Y}}
\newcommand {\sul}        {\mathrm{SU(2)_L}}
\newcommand {\suc}        {\mathrm{SU(3)_C}}
\newcommand {\ul}         {\mathrm{U(1)_Y}}
\newcommand {\uem}        {\mathrm{U(1)_{em}}}
\newcommand {\sigmabar}   {\overline{\sigma}}
\newcommand {\gmunu}      {g^{\mu \nu}}
\newcommand {\munu}       {{\mu \nu}}
\newcommand {\obra}       {\langle 0 |}
\newcommand {\oket}       {| 0 \rangle}
\newcommand {\bra}        {\langle}
\newcommand {\ket}        {\rangle}
%
\newcommand {\LL}         {L^{\alpha}_{\mathrm L}}
\newcommand {\LLd}        {L^{\dagger \alpha}_{\mathrm L}}
\newcommand {\lL}         {\ell^{\alpha}_{\mathrm L}}
\newcommand {\lLd}        {\ell^{\dagger \alpha}_{\mathrm L}}
\newcommand {\ld}         {\ell^{\dagger \alpha}}
\newcommand {\lb}         {\overline{\ell}^{\alpha}}
\newcommand {\lR}         {\ell^{\alpha}_{\mathrm R}}
\newcommand {\lRd}        {\ell^{\dagger \alpha}_{\mathrm R}}
\newcommand {\nuL}        {\nu^{\alpha}_{\mathrm L}}
\newcommand {\nuLb}       {\overline{\nu}^{\alpha}_{\mathrm L}}
\newcommand {\nub}        {\overline{\nu}^{\alpha}}
\newcommand {\lept}       {\ell^\alpha}
\newcommand {\neut}       {\nu^{\alpha}}
\newcommand {\nuLd}       {\nu^{\dagger \alpha}_{\mathrm L}}
\newcommand {\Phid}       {\Phi^\dagger}
%
\newcommand {\up}         {u^{\alpha}}
\newcommand {\ub}         {\overline{u}^{\alpha}}
\newcommand {\down}       {d^{\alpha}}
\newcommand {\db}         {\overline{d}^{\alpha}}
\newcommand {\QL}         {Q^{\alpha}_{\mathrm L}}
\newcommand {\QLd}        {Q^{\dagger \alpha}_{\mathrm L}}
\newcommand {\UL}         {U^{\alpha}_{\mathrm L}}
\newcommand {\ULd}        {U^{\dagger \alpha}_{\mathrm L}}
\newcommand {\UR}         {U^{\alpha}_{\mathrm R}}
\newcommand {\URd}        {U^{\dagger \alpha}_{\mathrm R}}
\newcommand {\DL}         {D^{\alpha}_{\mathrm L}}
\newcommand {\DLd}        {D^{\dagger \alpha}_{\mathrm L}}
\newcommand {\DR}         {D^{\alpha}_{\mathrm R}}
\newcommand {\DRd}        {D^{\dagger \alpha}_{\mathrm R}}
\newcommand {\bfell}      {\ell\kern-0.4em
                           \ell\kern-0.4em
                           \ell\kern-0.4em
                           \ell }
\newcommand {\obfell}     {\overline{\ell}\kern-0.4em
                           \overline{\ell}\kern-0.4em
                           \overline{\ell}\kern-0.4em
                           \overline{\ell}}
\newcommand {\bfH}      {\; {\cal H}\kern-0.5em \kern-0.4em
                           {\cal H}\kern-0.5em \kern-0.4em
                           {\cal H}\kern0.1em }
\newcommand {\obfH}     {\; \overline{\cal H}\kern-0.5em \kern-0.4em 
                           \overline{\cal H}\kern-0.5em \kern-0.4em 
                           \overline{\cal H}\kern0.1em }
%
\def \b             {{\mathrm b}}
\def \t             {{\mathrm t}}
\def \charm         {{\mathrm c}}
\def \d             {{\mathrm d}}
\def \u             {{\mathrm u}}
\def \e             {{\mathrm e}}
\def \q             {{\mathrm q}}
\def \g             {{\mathrm g}}
\def \p             {{\mathrm p}}
\def \s             {{\mathrm s}}
\def \y             {{\mathrm y}}
\def \n             {{\mathrm n}}
\def \l             {\ell} 
\def \f             {{f}} 
\def \D             {{\mathrm D}}
\def \K             {{\mathrm K}}
\def \Z             {{\mathrm Z}}
\def \W             {{\mathrm W}}
\def \S             {{\mathrm S}}
\def \N             {{\mathrm N}}
\def \L             {{\mathrm L}}
\def \R             {{\mathrm R}}
%
\newcommand {\dm}         {\Delta m}
\newcommand {\dM}         {\Delta M}
\newcommand {\ldm}        {\mbox{``low $\dm$''}}
\newcommand {\hdm}        {\mbox{``high $\dm$''}}
\newcommand {\nnc}        {{\overline{\mathrm N}_{95}}}
\newcommand {\snc}        {{\overline{\sigma}_{95}}}
\newcommand {\susy}       {{supersymmetry}}
\newcommand {\susyc}      {{supersymmetric}}
\newcommand {\aj}         {\mbox{\sf AJ}}
\newcommand {\ajl}        {\mbox{\sf AJL}}
\newcommand {\llh}        {\mbox{\sf LLH}}
%
\newcommand {\rpc}     {{\rm RPC}}
\newcommand {\rpv}     {{\rm RPV}}
\newcommand {\sfe}     {{\tilde{f}}}
\newcommand {\sfL}     {{\tilde{f}_{\mathrm L}}}
\newcommand {\sfR}     {{\tilde{f}_{\mathrm R}}}
\newcommand {\sfone}   {{\tilde{f}_{1}}}
\newcommand {\sftwo}   {{\tilde{f}_{2}}}
\newcommand {\sneu}    {{\tilde{\nu}}}
\newcommand {\wino}    {{\mathrm{\widetilde{W}}}}
\newcommand {\bino}    {{\mathrm{\widetilde{B}}}}
\newcommand {\se}      {{\mathrm{\tilde{e}}}}
\newcommand {\seR}     {{\mathrm{\tilde{e}_{R}}}}
\newcommand {\seL}     {{\mathrm{\tilde{e}_{L}}}}
\newcommand {\st}      {{\mathrm{\tilde{\tau}}}}
\newcommand {\stR}     {{\mathrm{\tilde{\tau}_{R}}}}
\newcommand {\stL}     {{\mathrm{\tilde{\tau}_{L}}}}
\newcommand {\stone}   {{\mathrm{\tilde{\tau}_{1}}}}
\newcommand {\sttwo}   {{\mathrm{\tilde{\tau}_{2}}}}
\newcommand {\sm}      {{\mathrm{\tilde{\mu}}}}
\newcommand {\smR}     {{\mathrm{\tilde{\mu}_{R}}}}
\newcommand {\smL}     {{\mathrm{\tilde{\mu}_{L}}}}
\newcommand {\Sup}     {{\mathrm{\tilde{u}}}}
\newcommand {\suR}     {{\mathrm{\tilde{u}_{R}}}}
\newcommand {\suL}     {{\mathrm{\tilde{u}_{L}}}}
\newcommand {\sdo}     {{\mathrm{\tilde{d}}}}
\newcommand {\sdR}     {{\mathrm{\tilde{d}_{R}}}}
\newcommand {\sdL}     {{\mathrm{\tilde{d}_{L}}}}
\newcommand {\sch}     {{\mathrm{\tilde{c}}}}
\newcommand {\scR}     {{\mathrm{\tilde{c}_{R}}}}
\newcommand {\scL}     {{\mathrm{\tilde{c}_{L}}}}
\newcommand {\sst}     {{\mathrm{\tilde{s}}}}
\newcommand {\ssR}     {{\mathrm{\tilde{s}_{R}}}}
\newcommand {\ssL}     {{\mathrm{\tilde{s}_{L}}}}
\newcommand {\stopR}   {{\tilde{\mathrm{t}}_{R}}}
\newcommand {\stopL}   {{\tilde{\mathrm{t}}_{L}}}
\newcommand {\stopone} {{\tilde{\mathrm{t}}_{1}}}
\newcommand {\stoptwo} {{\mathrm{\tilde{t}_{2}}}}
\newcommand {\sto}     {{\tilde{\mathrm{t}}}}
\newcommand {\SQ}      {{\mathrm{\widetilde{Q}}}}
\newcommand {\STO}     {{\mathrm{\widetilde{T}}}}
\newcommand {\glu}     {{\mathrm{\tilde{g}}}}
\newcommand {\sbotR}   {{\mathrm{\tilde{b}_{R}}}}
\newcommand {\sbotL}   {{\mathrm{\tilde{b}_{L}}}}
\newcommand {\sbotone} {{\mathrm{\tilde{b}_{1}}}}
\newcommand {\sbottwo} {{\mathrm{\tilde{b}_{2}}}}
\newcommand {\sbot}    {{\tilde{\mathrm{b}}}}
\newcommand {\squa}    {{\tilde{\mathrm{q}}}}
\newcommand {\squal}   {{\tilde{\mathrm{q}}_{\rm L}}}
\newcommand {\squar}   {{\tilde{\mathrm{q}}_{\rm R}}}
\newcommand {\sqL}     {{\tilde{\mathrm{q}}_{\rm L}}}
\newcommand {\sqR}     {{\tilde{\mathrm{q}}_{\rm R}}}
\newcommand {\snu}     {{\tilde{\nu}}}
\newcommand {\snue}    {{\tilde{\nu}_{\mathrm e}}}
\newcommand {\snum}    {{\tilde{\nu}_{\mu}}}
\newcommand {\snut}    {{\tilde{\nu}_{\tau}}}
\newcommand {\neu}     {{\chi}}
\newcommand {\chap}    {{\chi^+}}
\newcommand {\cham}    {{\chi^-}}
\newcommand {\chapm}   {{\chi^\pm}}

%
\newcommand {\thstop} {\mathrm{\theta_{\tilde{t}}}}
\newcommand {\thsbot} {\mathrm{\theta_{\tilde{b}}}}
\newcommand {\thsqua} {\mathrm{\theta_{\tilde{q}}}}
\newcommand {\Mcha}{M_{\chi^\pm}}
\newcommand {\Mchi}{M_\chi}
\newcommand {\Msnu}{M_{\tilde{\nu}}}
\newcommand {\tanb}{\tan\beta}
%

%
\newcommand {\rb}    {{\rm R_{\b}}}
\newcommand {\qq}    {{\q \overline{\q}}}
\newcommand {\bb}    {{\b \overline{\b}}}
\newcommand {\ff}    {{\f \bar{\f}}}
\newcommand {\el}    {{\e ^+}}
\newcommand {\po}    {{\e ^-}}
\newcommand {\ee}    {{\e ^+ \e ^-}}
\newcommand {\fbody} {{\sto \to \b \chi {\rm f \bar{f}'}}}
\newcommand {\gaga}  {\gamma\gamma}
\newcommand {\ggqq}  {\gamma\gamma \rightarrow \q\overline{\q}}
\newcommand {\ggtt}  {\gamma\gamma \rightarrow \tau^{+}\tau^{-}}
\newcommand {\qqg}   {\q\overline{\q}\gamma}
\newcommand {\ttg}   {\tau^{+}\tau^{-}\gamma}
\newcommand {\wenu}  {{\rm We\nu_\e}}
\newcommand {\gsZ}   {\gamma^\star\mathrm{Z}}
\newcommand {\ggh}   {\gamma\gamma\rightarrow{\mathrm{hadrons}}}
\newcommand {\ZZg}   {\mathrm ZZ^{*}/\gamma^{*}}
%
\newcommand {\zo}      {{z_0}}
\newcommand {\ip}      {{d_0}}
\newcommand {\thr}     {{{\rm thrust}}}
\newcommand {\athr}    {{\hat{\rm a}_{\rm thrust}}}
\newcommand {\ththr}   {{\theta_{\rm thrust}}}
\newcommand {\acol}    {{\Phi_{\rm acol}}}
\newcommand {\acop}    {{\Phi_{\rm acop}}}
\newcommand {\acopt}   {{\Phi_{\rm acop_T}}}
\newcommand {\thpoint} {\theta_{\rm point}}
\newcommand {\thscat}  {\theta_{\rm scat}}
\newcommand {\etwelve} {E_{12\gradi}}
\newcommand {\ethirty} {E_{30\gradi}}
\newcommand {\eiso}[1] {E^{\, \triangleleft 30\gradi}_{#1}}
\newcommand {\phimiss} {{\phi_{\vec{p}_{\rm miss}}}}
\newcommand {\ewedge}  {E(\phi_{\vec{p}_{\rm miss}}\pm 15\gradi)}
\newcommand {\evis}    {E_{\rm vis}}
\newcommand {\etot}    {E_{\rm vis}}
\newcommand {\emis}    {E_{\rm miss}}
\newcommand {\mvis}    {M_{\rm vis}}
\newcommand {\mtot}    {M_{\rm vis}}
\newcommand {\mmis}    {M_{\rm miss}}
\newcommand {\mhad}    {M^{\rm ex \, \ell_1}_{\rm vis}}
\newcommand {\mhadtwo} {M^{\rm ex \, \ell_1\ell_2}_{\rm vis}}
\newcommand {\ehad}    {E^{\rm NH}_{\rm vis}}
\newcommand {\epho}    {E^{\gamma}_{\rm vis}}
\newcommand {\echa}    {E^{\rm ch}_{\rm vis}}
\newcommand {\nch}     {{N_{\rm ch}}}
\newcommand {\elept}   {E_{\rm lept}}
\newcommand {\elepone} {E_{\ell _1}}
\newcommand {\eleptwo} {E_{\ell _2}}
\newcommand {\pvis}    {{\vec{p}_{\rm vis}}}
\newcommand {\pmis}    {{\vec{p}_{\rm miss}}}
\newcommand {\thmiss}  {{\theta_{\pmis}}}
\newcommand {\pt}      {{p_{\rm t}}}
\newcommand {\ptch}    {{p_{\rm t}^{\rm ch}}}
\newcommand {\pch}    {{p^{\rm ch}}}
\newcommand {\pz}      {{p_z}}
\newcommand {\ptnoNH}  {{p_{\rm t}^{\rm ex \, NH}}}
\newcommand {\puds}    {{P_{\rm uds}}}
%
%
\newcommand{\brchal}{\cal{B}($\PCha \rightarrow \ell\nu\PChi\ $)}
\newcommand{\M}{M_{2}}
\newcommand{\Mp}{M_{2}}
\newcommand{\sigbg}{\sigma_{\mathrm{bg}}}
\newcommand{\ww}   {\mathrm {WW}}
\newcommand{\zz}   {\mathrm Z\gamma^{*}}
\newcommand{\ewnu} {\mathrm{eW}\nu}
\newcommand{\eez}  {\mathrm {eeZ}}
\newcommand{\gagall}{{\gamma\gamma\rightarrow \ell\ell }}
\newcommand{\Pstaup}{{\widetilde{\tau}_{1}}}
\newcommand{\Pstaul}{{\widetilde{\tau}_{L}}}
\newcommand{\Pstaur}{{\widetilde{\tau}_{R}}}
\newcommand{\mzero}{m_{0}}
\newcommand{\msnu}{M_{\tilde{\nu}}}
\newcommand{\mcha}{M_{\chi^{\pm}}}
\newcommand{\mchi}{M_{\chi}}
\newcommand{\mstau}{M_{{\widetilde{\tau}_{1}}}}
\newcommand{\atau}{A_{\tau}}
\newcommand{\chsnu}{\PCha \rightarrow \ell \tilde{\nu}}
\newcommand{\chstau}{\PCha \rightarrow \tilde{\tau}_{1}\nu}
\newcommand{\chlep}{\PCha \rightarrow \ell\nu\chi}
\newcommand{\Tcsq}{\mathrm{TeV}/c^2}
\newcommand{\nobs}{N_{\mathrm{obs}}}
\newcommand{\nlim}{N_{\mathrm{lim}}}
\newcommand{\Brl}{\cal{B}_{\ell}}
\newcommand{\leff} {\mathcal{L}_{\mathrm{eff}}}
\newcommand{\dedx}{{\mathrm{d}}E/{\mathrm{d}}x}
\newcommand{\chtau}{\PCha \rightarrow \tau\nu\chi}
\newcommand{\ssqtw}{\sin^{2}\theta_{\mathrm W}}
\newcommand{\nnz}{{\mathrm \nu\bar{\nu}Z}}
\def \ggll    {\gamma\gamma \rightarrow \ell^{+}{\ell}^{-}}
\def \tautau  {\mathrm \tau^{+}\tau^{-}}
\def \ffg  {f\bar{f}(\gamma)}
\def \lll   {\ell^{+}{\ell}^{-}}
\def \ww   {\mathrm WW}
\def \zz   {\mathrm Z\gamma^{*}}
\def \znn  {\mathrm Z\nu\nu}
\def \zee  {\mathrm Zee}
\def \rts  {\sqrt{s}}
\def \mstop {m_{\tilde{\mathrm{t}}}}
\def \msnu  {m_{\tilde{\nu}}}
\def \elow   {E_{12^{\circ}}}
\def \gev    { \, \mathrm{GeV}/\it{c}^{\mathrm{2}}}
\def \gvm    { \, \mathrm{GeV}/\it{c}}
\def \mx     {M_{\mathrm{eff}}} 
\newcommand{\neutr}{\chi}


\def \Zcc           {\Z \to \charm \bar{\charm} }
\def \Zbb           {\Z \to \b \bar{\b} }
\def \decDS         {\D^{*+} \to \D^0 \pi^+}
\def \decsDS        {\D^{*+} \to \D^0 \pi^+_s}
\def \deckp         {\D^{0} \to \K^- \pi^+}
\def \deckppp       {\D^{0} \to \K^- \pi^+ \pi^+ \pi^-}
\def \deckpp        {\D^{0} \to \K^- \pi^+ \pi^0}
\def \deckpS        {\D^{0} \to \K^- \pi^+ (\pi^0)}
\def \decskp        {\D^{*+} \to \pi^{+}_{s} \K^- \pi^+}
\def \decskppp      {\D^{*+} \to \pi^{+}_{s} \K^- \pi^+ \pi^+ \pi^-}
\def \decskpp       {\D^{*+} \to \pi^{+}_{s} \K^- \pi^+ \pi^0}
\def \decskpS       {\D^{*+} \to \pi^{+}_{s} \K^- \pi^+ (\pi^0)}
\def \epsc          {\varepsilon_{\charm}}
\def \epsb          {\varepsilon_{\b}}
\def \pctod         {P_{\charm \to \D^*}}
\def \pbtod         {P_{\b \to \D^*}}
\def \Gbb           {\Gamma_{\b\bar{\b}}}
\def \Gcc           {\Gamma_{\charm\bar{\charm}}}
\def \Gh            {\Gamma_{\mathrm h}}

\title{The CMS Si-Strip Tracker}

\author{G.~Sguazzoni\footnote{\uppercase{o}n behalf of the
\uppercase{CMS} \uppercase{T}racker \uppercase{C}ollaboration.}}

\address{INFN Sezione di Pisa,\\
I-56127, Pisa (PI), Italy\\
E-mail: sguazzoni@pi.infn.it}

\maketitle

\abstracts{
The CMS experiment at LHC features the largest Silicon Strip Tracker
(SST) ever build. This device will be immersed in a 4T magnetic field
in conjunction with a Pixel system allowing the momentum
reconstruction of charged particles and the heavy-flavor tagging
despite the hostile radiation environment. The impact of the operating
conditions and the physics requirements on the SST layout and design
choices is discussed and the expected performances are reviewed. 
}

\section{Requirements and Implications}

The Compact Muon Solenoid (CMS) experiment will observe pp collision
at $14\TeV$ of center-of-mass energy with a luminosity of 
$10^{34}\cm^{-2}\s^{-1}$. The bunch crossing frequency is
$1/25\ns=40\MHz$ and $\sim20$ minimum bias interactions per bunch
crossing are expected, resulting in $\sim2000$ charged tracks per
event and a hadron flux up to $10^{14}\cm^{-2}\y^{-1}$ close to the
interaction region. 

\begin{figure}[t]
\centerline{\epsfxsize=9cm\epsfbox{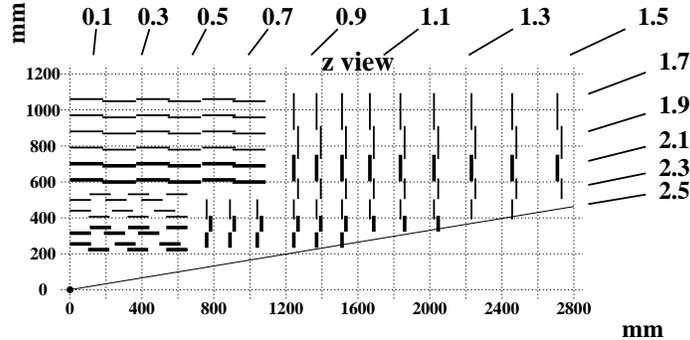}}
\caption{Sketch of one quadrant of the SST; thin and thick lines
  represent single- and double-sided modules,
  respectively.\label{fig:layout}} 
\end{figure}

The physics requirements, a momentum resolution of $\sim 1-2\%P_{T}$ at
$\sim 100\GeVc$ 
and an impact parameter resolution of $\sim10-20\um$, have to be
obtained by instrumenting a cylindrical volume of 5.4m in length and 2.4m
in diameter immersed in a 4T axial magnetic field. The innermost region
($r<15\cm$) is occupied by a pixel detector, described
elsewhere\cite{DKotlinski}.
The CMS collaboration decided to instrument the remaining volume by
using only Silicon microstrip modules organized in 10 cylindrical
layers and 12 disks as sketched in Figure~\ref{fig:layout},
corresponding to more than $200\m^2$ of active surface. A detailed
description of the Silicon Strip Tracker (SST) and its expected
performances can be found elsewhere\cite{TkTDR}. Only a brief overview
is given here.

\section{The Tracker Layout}

The tracker is divided into four main subsystem: the innermost four
cylindrical layers make up the Inner Barrel (TIB); the
outermost six cylindrical layers define the Outer Barrel (TOB);
the $2\!\times\!3$ disks with $|z|$ between $\sim70\cm$ and
$\sim110\cm$ are the Inner Disks (TID), each organized in three rings;
the bigger $2\!\times\!9$ disks in the 
$|z|\gtrsim120\cm$ region, organized in four to seven rings, are the Endcaps
(TEC). 

Most SST modules are single sided; the ``barrel'' modules and the
``disk'' modules have the readout strips laying along the $z$
direction and the radial direction, respectively, allowing the
readout of the $r\phi$ coordinate; however some threedimensional
information is needed to separate tracks. A space point readout is
obtained by using a double-sided module, a back-to-back sandwich of a
$r\phi$ module and 
a special {\it stereo} module with the strips tilted by $100\mrad$.
Double-sided modules equip the two innermost layers of TIB and TOB,
the two innermost rings of TID, and the TEC rings with
$r\lesssim40\cm$ and with $60{\cm}\lesssim r\lesssim76\cm$ (see
Figure~\ref{fig:layout}).
This layout allows the SST to provide 8 to 14 measurements points for high
momentum tracks with $|\eta|<2.5$. On average about half of them are
threedimensional points. 

\section{The modules}

The overall dimensions of the SST, the largest device of this kind
ever built, require such a huge number of basic elements to impose an
industrial approach to the design and the production. The SST modules
and their components share the same basic structure and design. Each
module consists of a carbon fiber frame that supports the silicon
detector and the readout electronics, hosted on a front-end hybrid.    

The modules have to comply with the following characteristics: a pitch
of the order of $\sim 100\um$ to ensure the target momentum resolution;
radiation resistance with no significant performance degradation for
the 10-years of LHC lifetime; high granularity in time (low
pile-up, i.e. $\sim25\ns$ shaping time) and in the space domain
(occupancy below $1\%$, i.e. a cell size of $\sim 1\cm^2$,
corresponding to a strip length of the order of $\sim10\cm$) to ensure
a robust and efficient pattern recognition.

All the SST detectors are standard p$^+$-on-n microstrip sensors with AC
readout and polysilicon bias resistor, produced by single-sided
lithographic process on 6''-wafer industrial lines, thus allowing an
effective cost reduction. The sensor radiation hardness relies upon several
features\cite{SBraibant}: special design details like
multi-guard rings, the constant width-over-pitch of 0.25, and the metal
over-hang (metal strips wider than 
underlying implants) effectively improve the breakdown voltage
behavior; the use of low-resistivity bulk 
($\sim1.5-8{\rm k}\Omega \cm$) allows the sensors to be depleted at a
manageable voltage over the entire LHC lifetime also after the
type inversion; an interstrip capacitance almost fluence independent
is obtained by using the $\bra100\ket$ lattice orientation that gives
a Si-SiO$_2$ interface of better quality.

Nevertheless, the SST operating and storage temperature must be below
$-10^\circ$ to keep under control the radiation-induced increase of
the leakage current and to freeze-out the unwanted reverse annealing.

\begin{figure}[ht]
  \begin{center}
    \begin{picture}(0,0)
      \put(0,-10){\mbox{(a)}}
    \end{picture}
    \epsfig{file=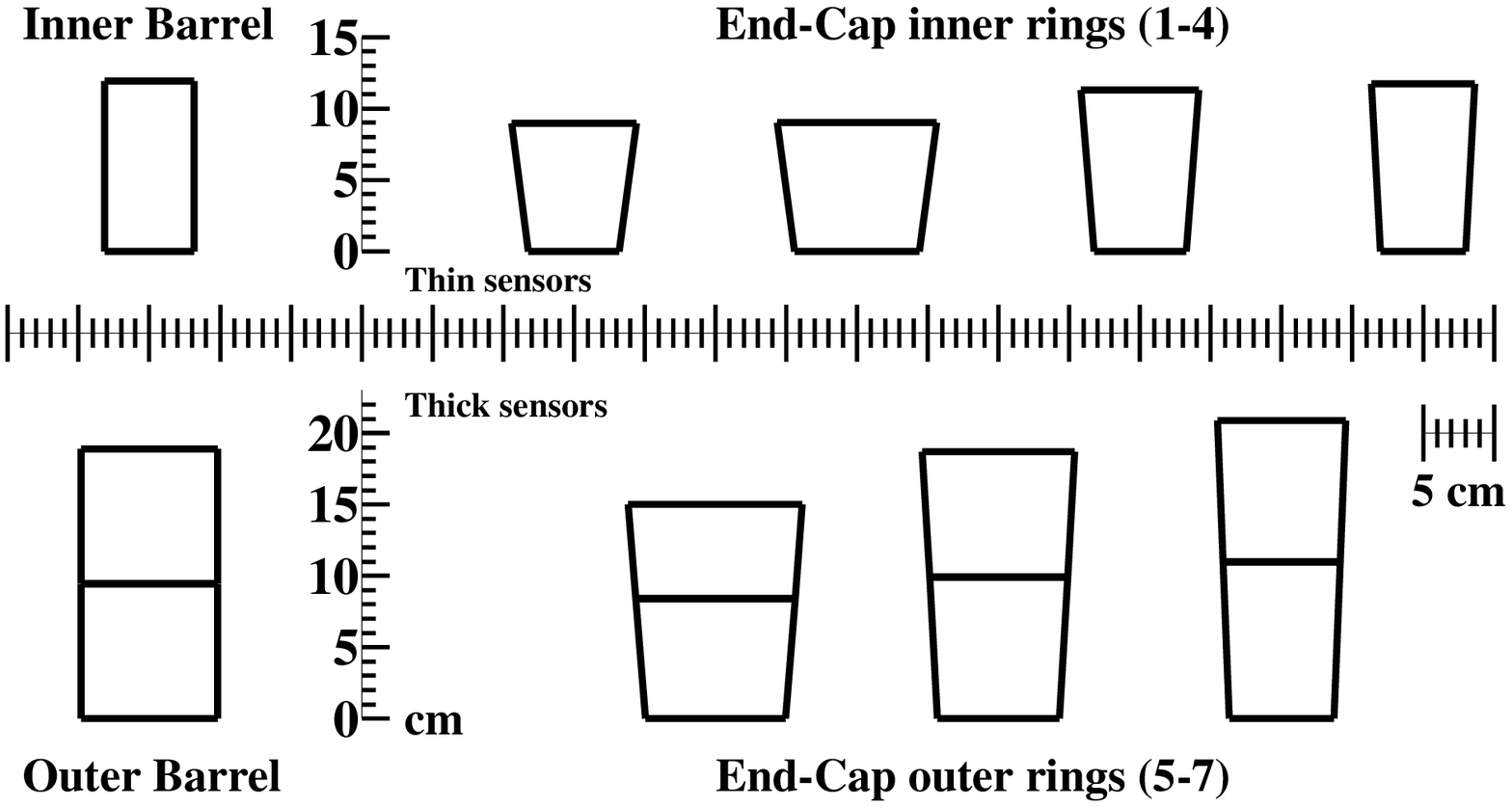,height=2.6cm}
    \hskip 0.04\textwidth
    \begin{picture}(0,0)
      \put(0,-10){\mbox{(b)}}
    \end{picture}
    \epsfig{file=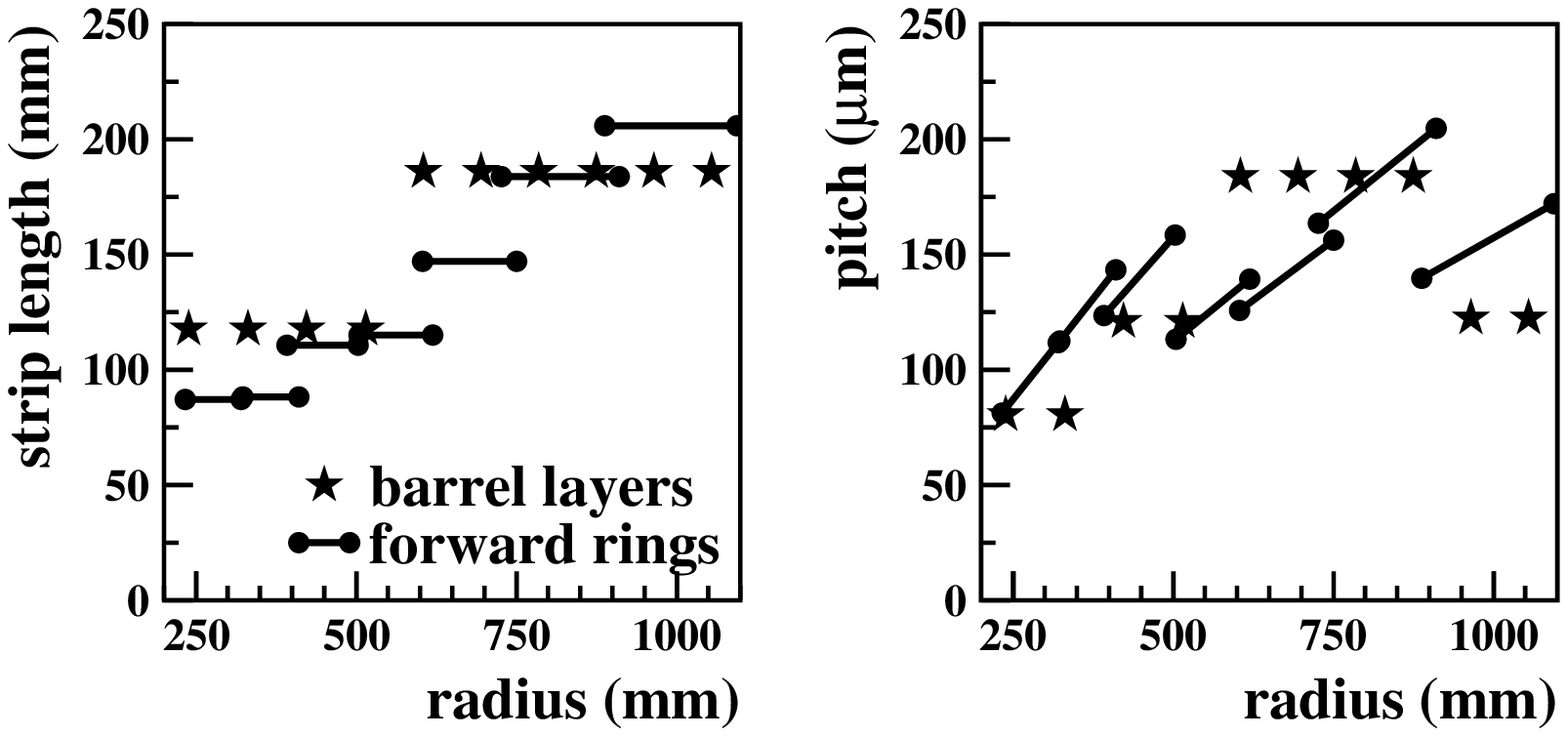,height=2.6cm}
  \end{center}
\caption{SST detectors shapes and dimensions (a) and strip length and
  pitch evolution with the radius (b).\label{fig:sens}}
\end{figure}

A key design aspect of the SST to limit the costs and the number of
readout channels relies upon the scaling of track multiplicity and
radiation effects 
with the radius. Occupancy requirement imposes sensors within
$\sim60\cm$ in radius 
to have a pitch of $\sim80 - 120\um$ and a strip length of $\sim
10\cm$. To limit the number of readout channels this requirement is 
relaxed at $r\gtrsim60\cm$ choosing a pitch of $\sim120
- 200\um$ and a strip length of $\sim 20\cm$, obtained by
daisy-chaining two single sensors. Similarly, $320\um$ thick sensors
with $\sim1.5-3.2{\rm k}\Omega \cm$ resistivity are needed for
$r\lesssim60\cm$ to have a 
comfortable depletion voltage after the irradiation but, for
$r\gtrsim60\cm$, less expensive $500\um$ thick sensors with
$\sim4-8{\rm k}\Omega \cm$ resistivity can be safely chosen, also 
to compensate the increase of noise due to the larger strip length.  
The various shapes and dimensions of the SST detectors are shown in
Figure~\ref{fig:sens}(a) and the evolution of strip length and pitch
is shown in Figure~\ref{fig:sens}(b).

The front-end hybrid is made of a kapton circuit (that also integrates the
connection tail) laminated onto a ceramic rigidifier. It hosts four or six 
APV25s, the SST front-end chips, the additional ASICs (APVMUX, PLL and
DCU) and the pitch adapter that allows the APV25 channels' pitch to
match the sensors' one. 

The APV25 has been designed in $0.25\um$ CMOS technology for
low-noise and fast signal readout in high radiation environment. 
It has 128 charge-sensitive amplifying channels with $\sim50\ns$ shaping
time multiplexed into a single readout line. A deconvolution filter is
implemented to reduce the time resolution to $\sim25\ns$. A
pipeline buffer stores analog samples for 192 bunch crossings,
corresponding to $\sim 4.8\mu\s$, to match the LVL1 trigger
requirements. 
Further details on the front-end hybrids and a description of the
downstream readout chain can be found elsewhere\cite{KKlein}.

The assembly of the $\sim 15000$ modules needed to complete the entire
SST is performed by means of semi-automatic {\it gantries} featuring
pattern recognition for an accurate and reproducible placement of the
components with tolerances that are comparable with the intrinsic
resolution of the device ($\sim10-20\um$). Similarly the $\sim25$M
bonds required to connect sensors are made by automatic bonding machines.

Production is actually starting. A tight quality assurance procedure
is strictly enforced at each production/assembly step to check that all
components match the required specifications. These tests normally
include cooling cycle to reproduce the actual working condition.

\section{Mechanics and Material Budget}

The modules support structure is made up of carbon fiber and
honeycomb frames. It ensures a mechanical stability within $\sim
20\um$ and the $\lesssim 100-200\um$ absolute position accuracy needed by
track finding and alignment procedures\cite{SKoenig}. The
structure also hosts services like readout lines, power cables and
cooling pipes (to take out $\sim 1\W/{\rm module}$) and has been
designed to grant easy access and operation on all subelements. 

\begin{figure}[t]
  \begin{center}
    \begin{picture}(0,0)
      \put(-4,0){\mbox{(a)}}
    \end{picture}
    \epsfig{file=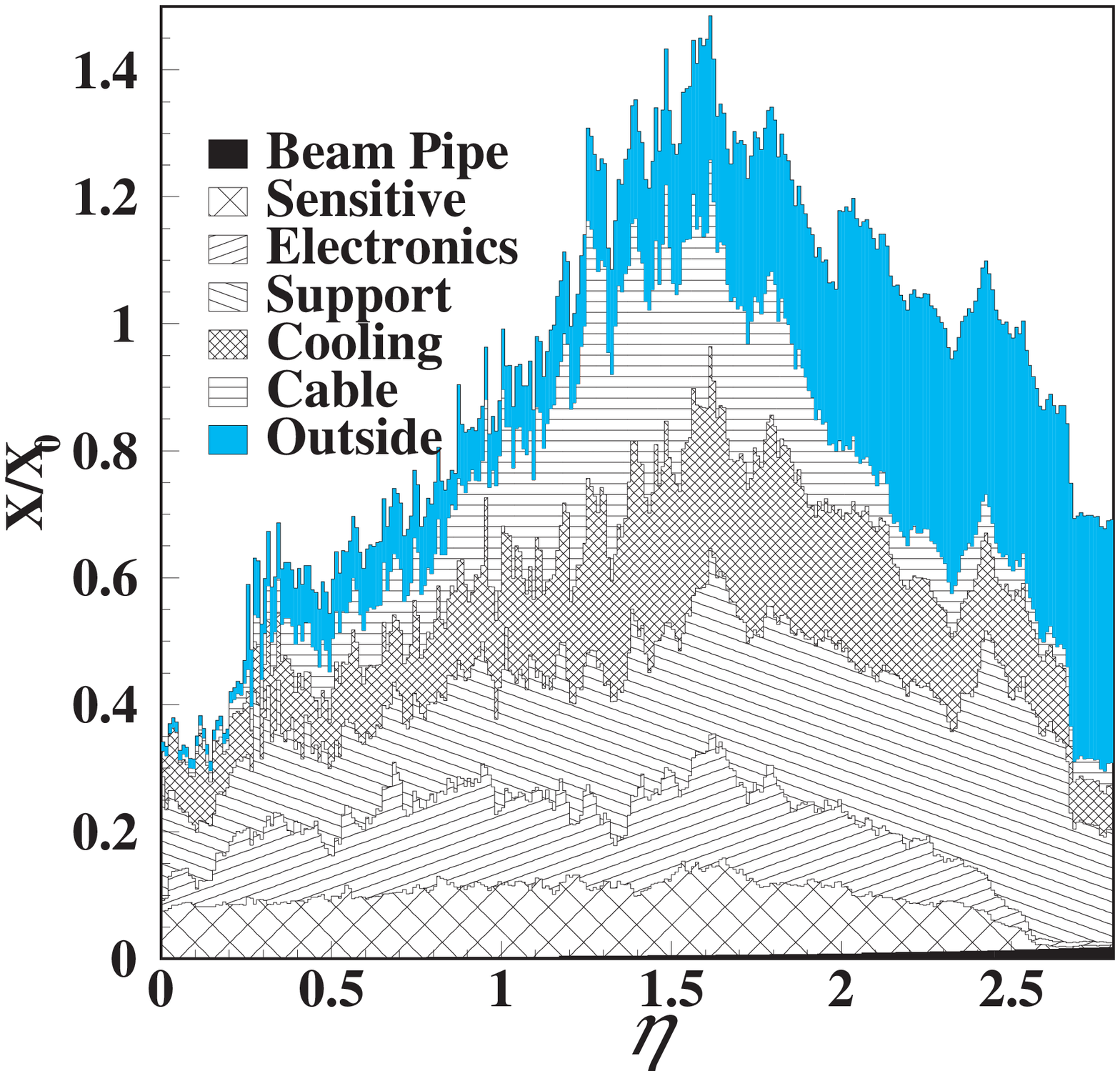,height=3.2cm}
    \hskip 0.03\textwidth
    \begin{picture}(0,0)
      \put(-2,0){\mbox{(b)}}
    \end{picture}
    \epsfig{file=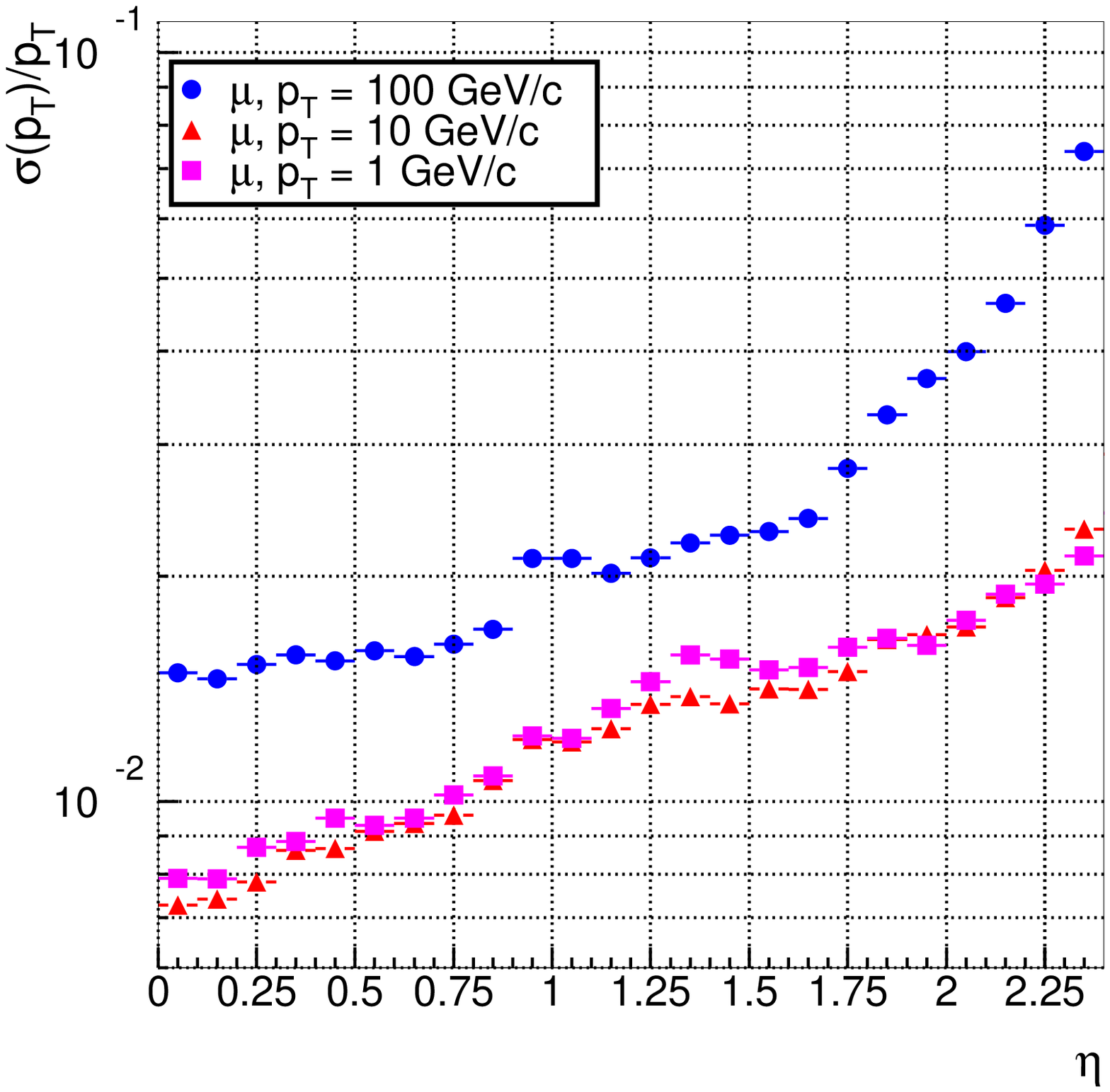,height=3.2cm}
    \hskip 0.03\textwidth
    \begin{picture}(0,0)
      \put(-2,0){\mbox{(c)}}
    \end{picture}
    \epsfig{file=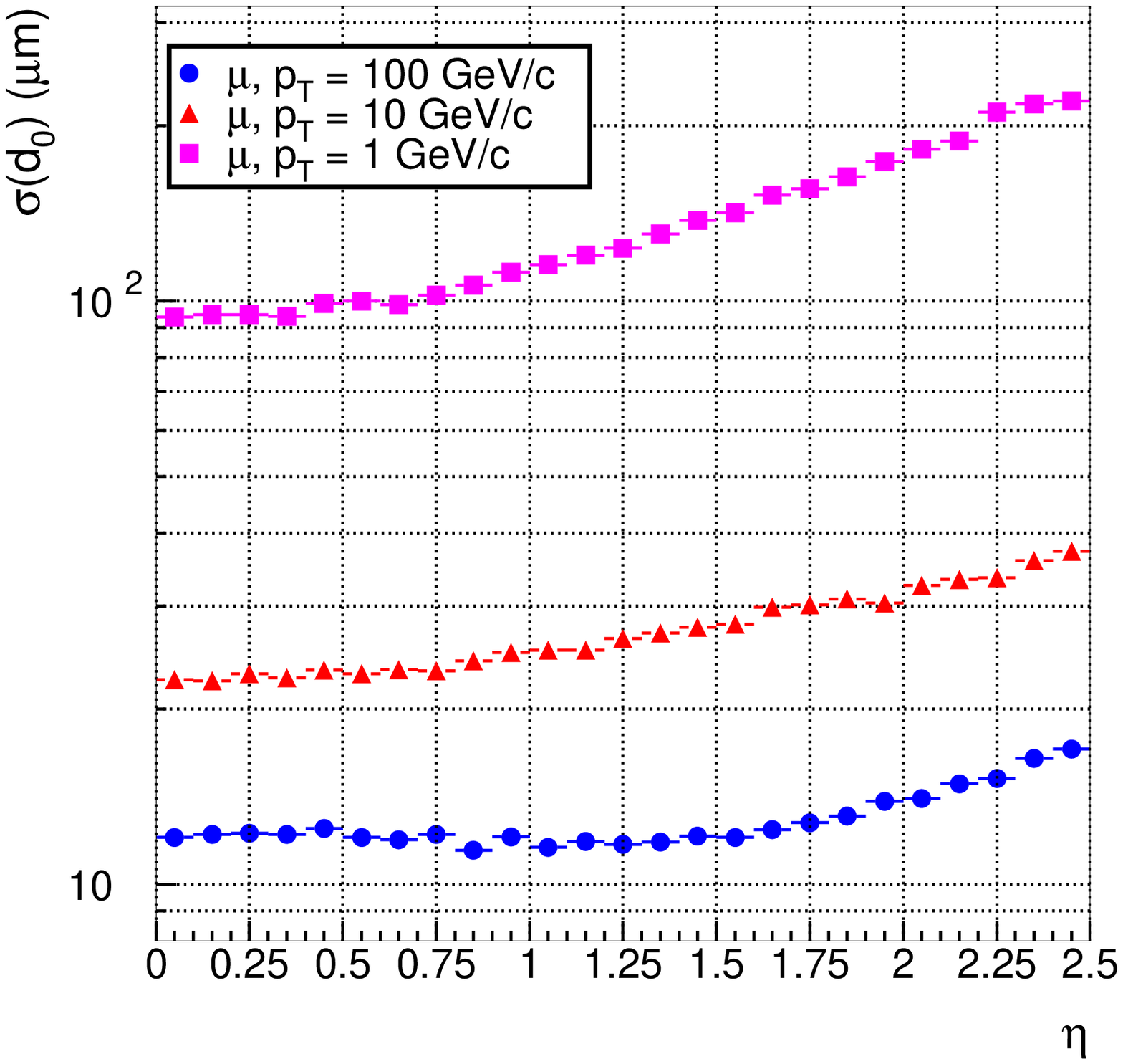,height=3.2cm}
  \end{center}
\caption{SST material budget in number of radiation lengths as a
  function of $\eta$ (a); $P_T$ resolution (b) and transverse impact
  parameter resolution (c) for muons of various momenta as a
  function of $\eta$.\label{fig:mat}} 
\end{figure}

Despite all optimization efforts to keep the SST as light as possible
the contribution of active material, structures, electronics and
services results into a substantial material budget. Figure~\ref{fig:mat}(a)
shows the result of the detailed GEANT4 simulation in terms of
radiation lengths as a function of pseudorapidity $\eta$.
\begin{figure}[b]
  \begin{center}
    \begin{picture}(0,0)
      \put(3,-1){\mbox{(a)}}
      \put(101,-1){\mbox{(b)}}
    \end{picture}
    \epsfig{file=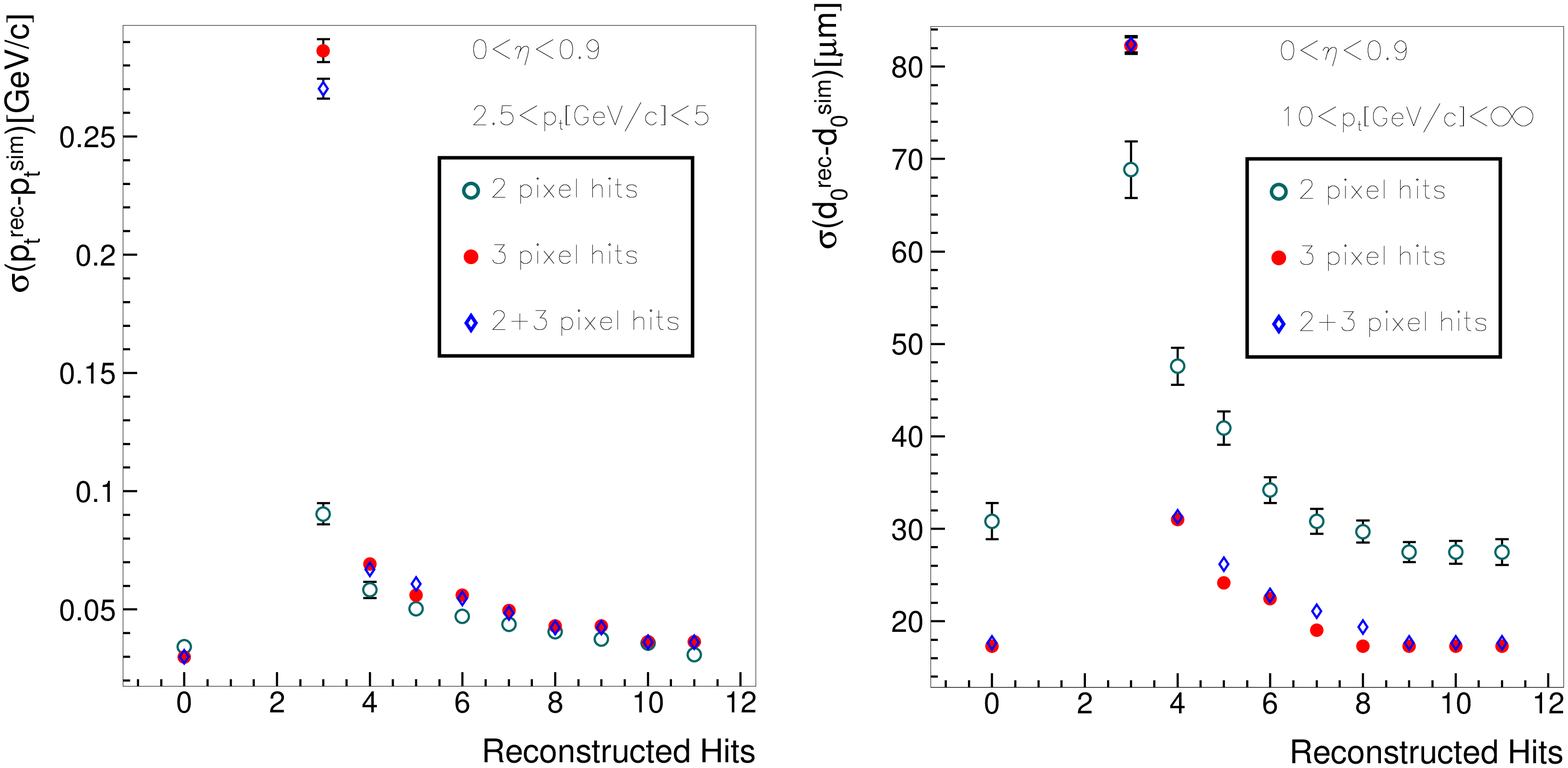,height=3.2cm}
    \hskip 0.01\textwidth
    \epsfig{file=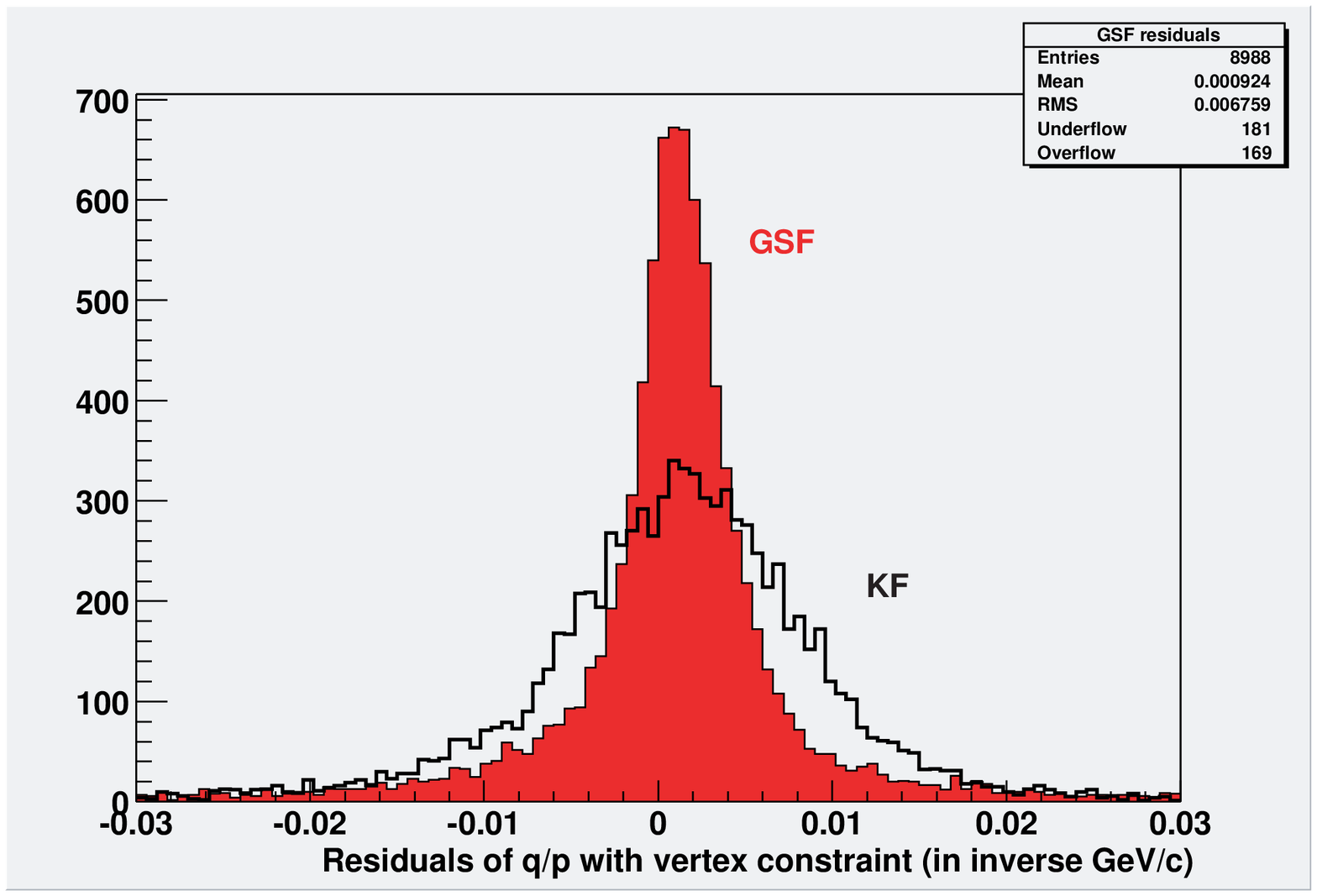,height=3.2cm, width=4.1cm}
    \begin{picture}(0,0)
      \put(-121,-1){\mbox{(c)}}
    \end{picture}
  \end{center}
\caption{$P_T$ resolution (a) and transverse IP resolution (b) as a
  function of the number of reconstructed SST hits, using also the Pixel
  detector in various configurations; the full tracker resolutions are
  the values at ``zero'' reconstructed hits. (c) Full simulation
  residuals of the 
  estimated curvature (q/p) with respect to the true value at the transverse
  impact point for the KF and the GSF (see text). A vertex constraint
  has been included.\label{fig:fast}}
\end{figure}

\section{Performance}

The $P_T$ resolution, shown for muons in Figure~\ref{fig:mat}(b), is
$\sim2\%$ or better for $P_{T} < 100 \GeVc$ and $|\eta|<1.7$; at
larger pseudorapidity the performance degrades because of 
the reduction of the lever arm. The transverse impact parameter (IP)
resolution, shown in Figure~\ref{fig:mat}(c), is $\lesssim20\um$ in
the entire $\eta$ range covered by the tracker, for muons of $P_T = 100
\GeVc$. Low momentum $P_T$ and IP resolutions are degraded by the
multiple scattering. The track reconstruction 
efficiency is close to 100\% for muons in most of the pseudorapidity
range, while it drops to $90\%-95\%$ for pions and tracks within jets,
mostly because of nuclear interactions. 

As shown in Figure~\ref{fig:fast}(a) and (b), sufficient track
reconstruction accuracy is achieved by using the pixel hits and a
reduced number of silicon strip hits, tipically four to six. Such
figures demonstrate the redundancy and the robustness of the SST
layout and the possibility to use fast tracking algorithms for trigger
applications\cite{DAQTDR}. 

The negative impact of the large material budget on the SST
performances can be reduced by designing reconstruction algorithms
that take into account the presence of material. As an
example, electrons suffer for large material-dependent
bremsstrahlung energy losses. A precise modeling of this effect has
been implemented in a special electron reconstruction
algorithm\cite{WAdam} by using the {\it Gaussian-sum Filter} (GSF), a
non-linear generalization of the standard {\it Kalman Filter}
(KF). The impressive gain in resolution of this special electron
reconstruction algorithm is shown in Figure~\ref{fig:fast}(c). 

\section{Conclusion}

The present silicon strip technology allows the CMS collaboration to
build a large scale tracker despite the difficult operating
environment of a high-luminosity, high-energy hadron
collider. Detailed simulation studies suggest that the SST
performances will lay within the physics requirements. The layout 
redundancy makes possible to use fast-tracking algorithms at trigger
level. The effects of the relatively large amount of material inside
the tracking volume can be taken into account by accurate modeling
and dedicated algorithms.

\end{document}